%% file: production_release_2_10_0.tex
\documentclass[11pt,epsfig,epsf]{article}
\usepackage{graphicx}

\usepackage{booktabs}

\usepackage{lineno}

\usepackage{authblk}

\usepackage{hyperref}

\usepackage{color}

\usepackage{cite}

\usepackage{textcomp} 
\input{preamble.tex}

\usepackage{titlesec}
\titleformat{\paragraph}
{\normalfont\normalsize\bfseries}{\theparagraph}{1em}{}
\titlespacing*{\paragraph}
{0pt}{3.25ex plus 1ex minus .2ex}{1.5ex plus .2ex}

\begin{document}

\title{GENIE Production Release 2.10.0}

\author[1]{M.~Alam}
\author[2,3]{C.~Andreopoulos}
\author[1]{M.~Athar}
\author[4]{A.~Bodek}
\author[5]{E.~Christy}
\author[4]{B.~Coopersmith}
\author[6]{S.~Dennis}
\author[7]{S.~Dytman}
\author[8]{H.~Gallagher}
\author[7]{N.~Geary}
\author[4,9]{T.~Golan}
\author[9]{R.~Hatcher}
\author[10]{K.~Hoshina}
\author[11]{J.~Liu}
\author[12]{K.~Mahn}
\author[4]{C.~Marshall}
\author[12]{J.~Morrison}
\author[13]{M.~Nirkko}
\author[14]{J.~Nowak}
\author[9]{G.~N.~Perdue}
\author[9]{J.~Yarba}
\affil[1]{\Aligarh}
\affil[2]{\Liverpool}
\affil[3]{\STFC}
\affil[4]{\Rochester}
\affil[5]{\Hampton}
\affil[6]{\Warwick}
\affil[7]{\Pittsburgh}
\affil[8]{\Tufts}
\affil[9]{\FNAL}
\affil[10]{\Wisconsin}
\affil[11]{\WM}
\affil[12]{\MSU}
\affil[13]{\UBern}
\affil[14]{\Lancaster}

\maketitle
%
%

\begin{abstract}
GENIE (Generates Events for Neutrino Interaction Experiments) is a neutrino Monte Carlo event generator that simulates the primary interaction of a neutrino with a nuclear target, along with the subsequent propagation of the reaction products through the nuclear medium.
It additionally contains libraries for fully-featured detector geometries and for managing various types of neutrino flux.
This note details recent updates to GENIE, in particular changes introduced into the newest production release, version 2.10.0.
\end{abstract}

\section{Introduction}

GENIE \cite{Andreopoulos:2009rq} is a neutrino event generator created to be the ``universal event generator'' discussed during the NUINT conference series. 
It was designed to provide an accessible, extensible framework with many convenience tools leveraging existing HEP software like \pythia \cite{Sjostrand:2006za}, ROOT \cite{Brun:1997pa}, and LHAPDF \cite{Whalley:2005nh}.

This note describes the changes in the GENIE generator code base between production release version 2.8.0 (and its subsequent patches) and the new production release, 2.10.0. 
The GENIE code is available through a publicly visible Subversion repository or via a source tar-file.
Details for both methods of accessing the code are available on the GENIE homepage at \url{http://genie.hepforge.org}.

GENIE 2.10.0 is a ``model introduction release.'' 
Broadly speaking, the GENIE collaboration releases two kinds of updates for the generator code - model introduction and physics tuning releases.
Model introduction releases aim to incorporate new models into the code base but they do not incorporate them into the default global physics model.
Physics tuning releases modify the default global physics model.
Note that these categories are not hard and fast rules - occasionally the global physics model will perform slightly differently after a bug is removed from the code in a model introduction release, and we will occasionally introduce new models in physics tuning releases.

\section{Modifications to Existing Models}

In this section we describe new models that can be utilized as options within the existing GENIE generators.      None of these 
are turned on by default (with the exception of the inclusion of the W/Z propagator terms in the DIS cross section), but can be
enabled by the user in the UserPhysicsOptions.xml file.  

\begin{figure}
  \centering
  \includegraphics[height=0.65\textheight]{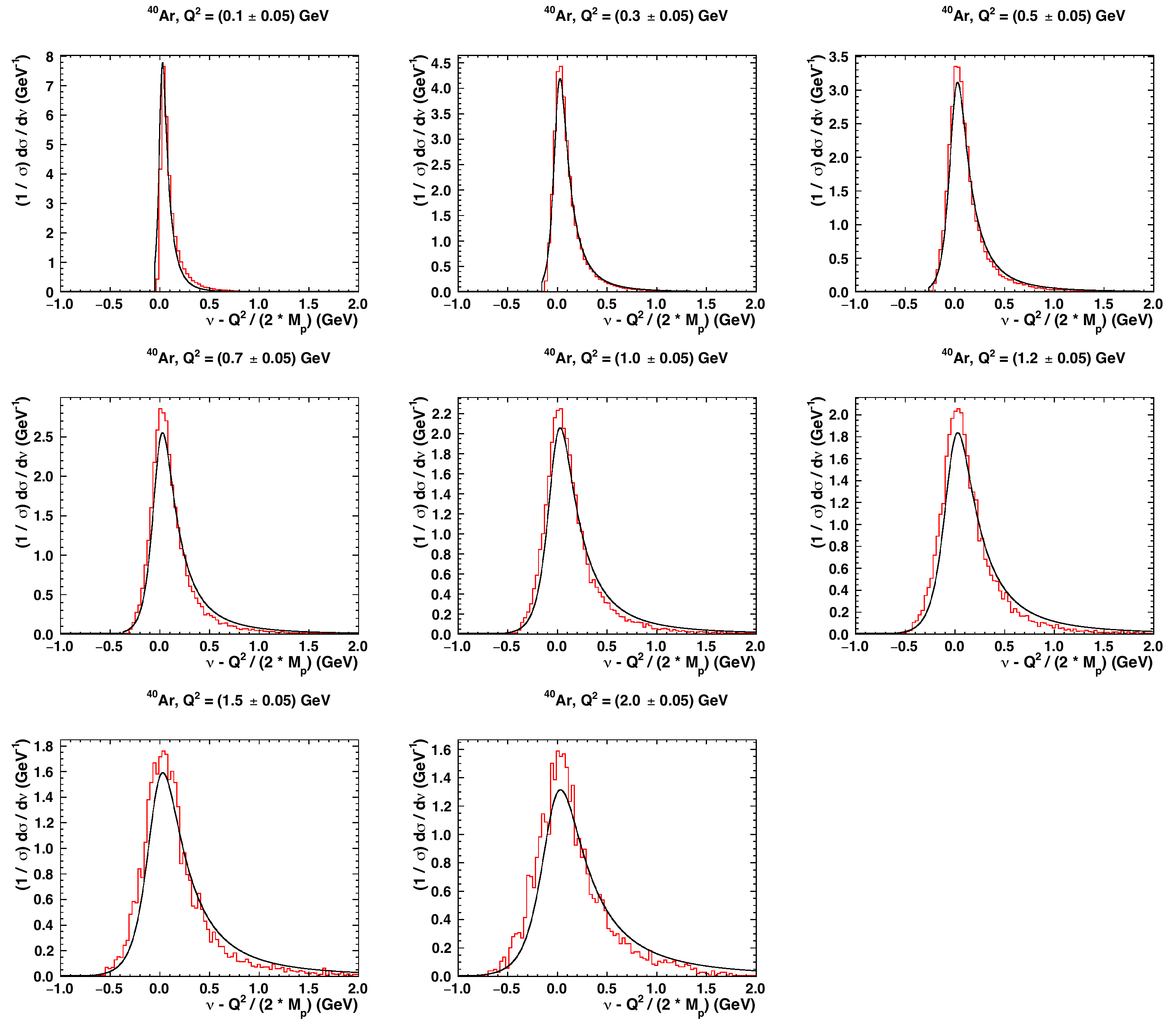}
  \caption{Comparisons between the $\psi'$ superscaling parameter as a function of $\nu - Q^2/\left(2 M_p\right)$, where $\nu$ is the energy transfer to the hadronic system during the neutrino scattering event, $Q^2$ is the four momentum transfer from the neutrino, $M_p$ is the target nucleon mass. 
The curve extracted from electron scattering data \cite{Bodek:2014pka} (the smooth black curve) is compared to the computation produced by GENIE (the red histogram) for different values of $Q^2$ on Argon-40.}
  \label{fig:superscalingargonplot}
\end{figure}

\subsection{Nuclear Models}

Quasi-elastic (QE) models based on Llewelyn Smith \cite{LlewellynSmith:1971zm} and the Fermi Gas Model \cite{PhysRevLett.26.445} are presently the standard in neutrino event generators because of their simplicity and wide applicability.  
They qualitatively describe inclusive electron scattering data, but struggle to explain exclusive electron and neutrino cross section results consistently over a wide range of energies and kinematics \cite{PhysRevD.81.092005, PhysRevD.88.032001, PhysRevLett.111.022501, PhysRevLett.111.022502, NOMADCCQE}.
One explanation for this is an incomplete description of nuclear effects, with nucleon-nucleon correllations such as Meson Exchange Currents expected to play a large role \cite{Donnelly:1999sw,Carlson:2001mp,Maieron:2009an,Martini:2009uj,Amaro:2010sd,Martini:2010ex,Nieves:2013fr}.
These modify both the nucleon momentum distribution and the shape of the cross section.
Two models were added to GENIE to capture these effects: a Transverse Enhancement model (TEM) \cite{transverseenahnce} and an Effective Spectral Function model \cite{Bodek:2014pka}.

\subsubsection{Transverse Enhancement}

In the TEM, nuclear effects like those expected from Meson Exchange Currents are modeled as $Q^2$ dependent modifications to the elastic proton and neutron magnetic form factors.
The exact $Q^2$ dependence is extracted from fits of the transverse quasi-elastic response function from electron scattering data.
This model by itself doesn't emit nucleons.  However, 1 or 2 nucleons are emitted through the effective spectral function.

\subsubsection{Effective Spectral Functions}

Nuclear models such as the Local Thomas Fermi Gas \cite{PhysRevC.79.034601}, global Fermi Gas \cite{PhysRev.184.1154,PhysRevLett.26.445,Smith1972605}, or Benhar-Fantoni Spectral Function \cite{PhysRevC.55.244,benfanspecfun2} provide different momentum distributions which changes the shape of the quasielastic cross section.
Final state interactions at the Feynman diagram level change the shape of the differential cross section in energy transfer, with an increase of strength in the tail of the distribution and an decrease in the peak.
These interactions are included in superscaling calculations \cite{PhysRevC.71.015501}. 
The Effective Spectral Function model in GENIE 2.10.0 is fitted to these model's predictions of 
$1/\sigma \times d\sigma/d\nu$, where $\nu$ is the energy transfer to the hadronic system during the scattering event, at various values of $Q^2$, where $Q^2$ is the four momentum transfer from the neutrino.
See Figure \ref{fig:superscalingargonplot} for a comparison of the superscaling model prediction to the predictions of implementation in GENIE.

\begin{figure}
  \centering
  \includegraphics[height=0.45\textwidth]{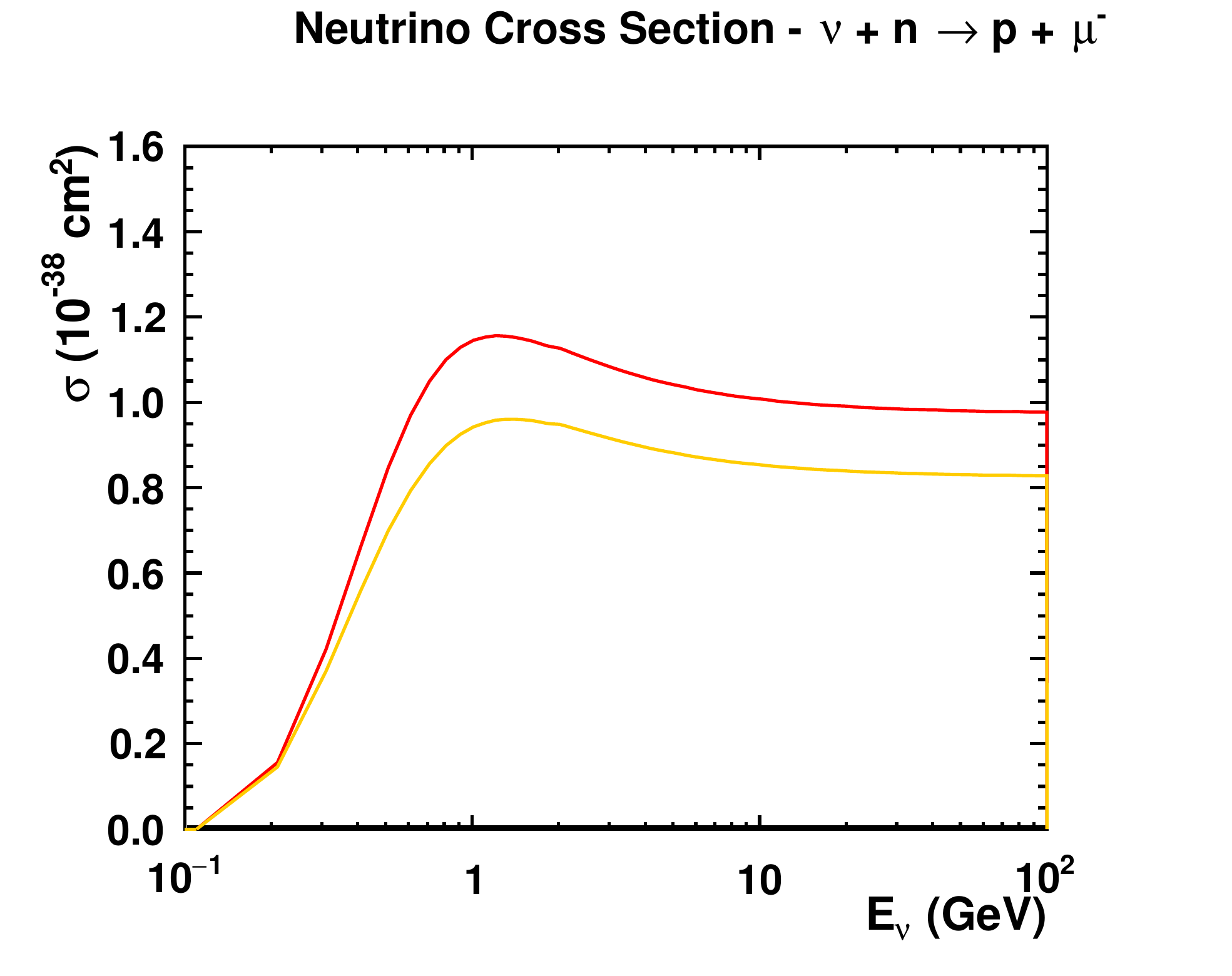}
  \caption{The charged-current quasielastic cross section for neutrinos with the default Llewelyn Smith model in orange and the new Effective Spectral Function model in red.
The cross section is computed on Carbon and then scaled by the number of neutrons.}
  \label{fig:effspecfunc_nu_xs_energy}
\end{figure}

\begin{figure}
  \centering
  \includegraphics[height=0.45\textwidth]{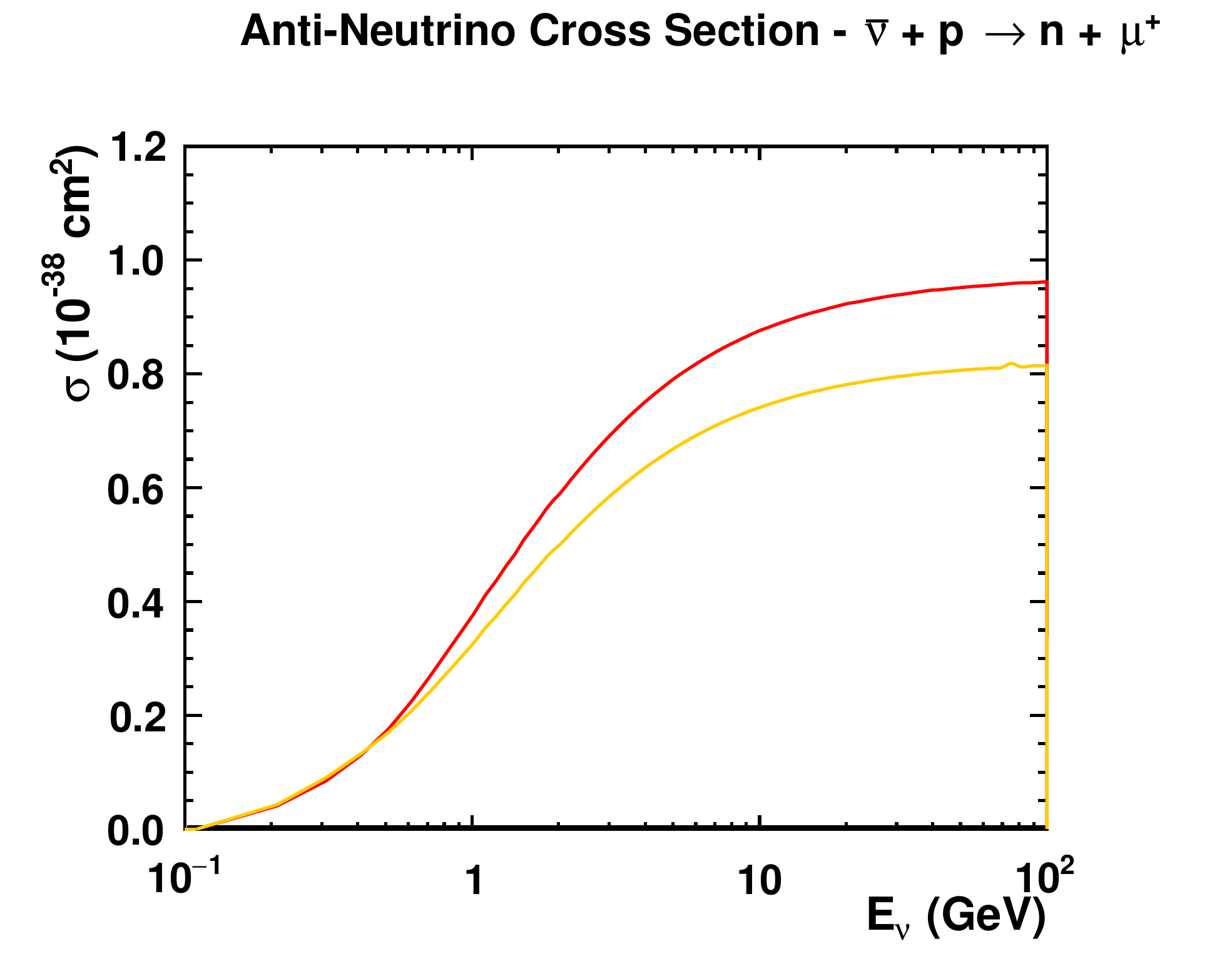}
  \caption{The charged-current quasielastic cross section for antineutrinos with the default Llewelyn Smith model in orange and the new Effective Spectral Function model in red.
The cross section is computed on Carbon and then scaled by the number of protons.}
  \label{fig:effspecfunc_nu_bar_xs_energy}
\end{figure}

The TE Model uses a modified transverse form factor to add to the strength to higher energy loss.
Used with the Effective Spectral Function as the nuclear model, the TEM effectively provides a Meson Exchange Currents.
Together, they give a complete prescription that is in agreement with a wide range of electron scattering data.
The EFS and TEM models may be activated independently if the user chooses and they are not active by default in GENIE 2.10.0.
Figures \ref{fig:effspecfunc_nu_xs_energy} and \ref{fig:effspecfunc_nu_bar_xs_energy} show the neutrino and antineutrino cross sections on Carbon as a function of energy with the EFS active (but TEM disabled) scaled by the number of protons or neutrons as appropriate.

\subsubsection{Activating Effective Spectral Functions and Transverse Enhancement}
\label{sec:activateeffspecfunc}

By default, GENIE uses the Relativistic Fermi Gas with the Bodek-Ritchie modifications \cite{PhysRevD.23.1070,PhysRevD.24.1400} and no Transverse Enhancement.
To change the model used, make edits to the configuration file found in \texttt{UserPhysicsOptions.xml}.
Activate Transverse Enhancement by changing the parameter \texttt{UseElFFTransverseEnhancement} to \texttt{true}.
Similarly, the Effective Spectral Functions is activated by changing the \texttt{NuclearModel} to \texttt{genie::EffectiveSF/Default}.
To activate both models simultaneously, make both of the changes described above.

\subsection{Hadronization Models}

\begin{figure}
  \centering
  \includegraphics[width=0.50\textwidth]{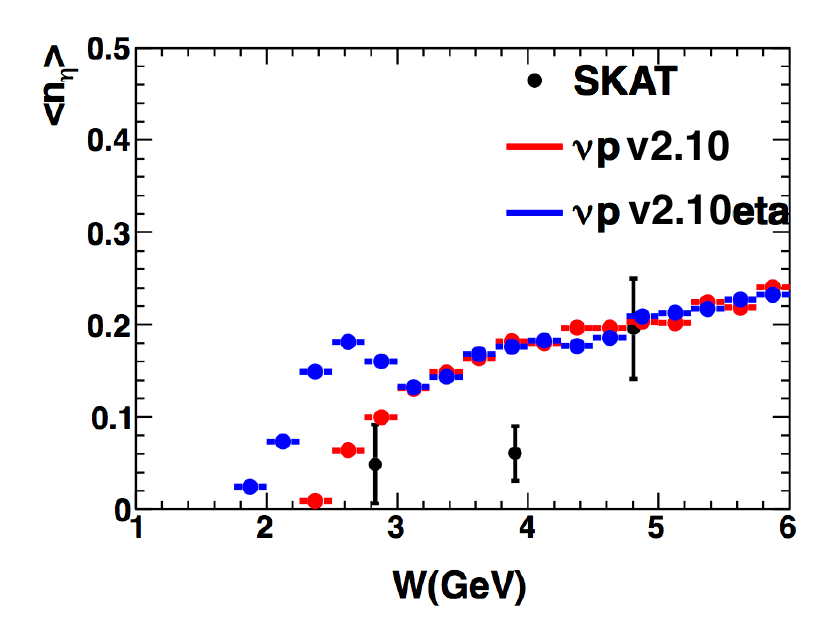}
  \caption{Eta production rate measurements from the SKAT experiment \cite{Agababyan:2008kx}, compared with the GENIE default prediction (red) and the
  GENIE prediction with eta production parameters set to large non-zero values (blue).
  \label{fig:etas}}
 \end{figure}

Eta mesons, like $\pi^0$s, have purely electromagnetic decays into photons and can therefore mimic electron neutrino appearance.
For this reason, their simulation is important for oscillation experiments.
Prior to this GENIE release, $\eta$ mesons were produced through two mechanisms, the decay of baryon resonances, and \pythia fragmentation.
The result is a kinematic gap over which $\eta$ mesons are not produced - non-resonant inelastic events with invariant masses too low to be fragmented by \pythia.
These events are handled by the KNO-based part of the AGKY model \cite{Yang:2009zx}.
In this model, mesons are produced in pairs with a net charge of zero, according to probabilities assigned via the {\tt KNO-Prob*} values in \texttt{UserPhysicsOptions.xml}.
Two new values have been added in this release, {\tt KNO-ProbPi0Eta} and {\tt KNO-ProbEtaEta}.
The ability to create $\eta$ mesons over all values of W makes possible background studies for oscillation experiments.
Both are currently set to zero in 2.10.0, but we expect that they will be tuned to non-zero values in the next GENIE physics release.
Figure~\ref{fig:etas} shows the effect of setting these parameters to a value of 0.1, with a corresponding decrease in other KNO-Prob values.
These are unrealistically large values, and are used here for the purposes of illustrating the kinematic range that is being affected by these parameters.

\subsection{Intranuclear Rescattering Models}

\begin{figure} [tp]
  \centering
  \includegraphics[height=0.25\textheight]{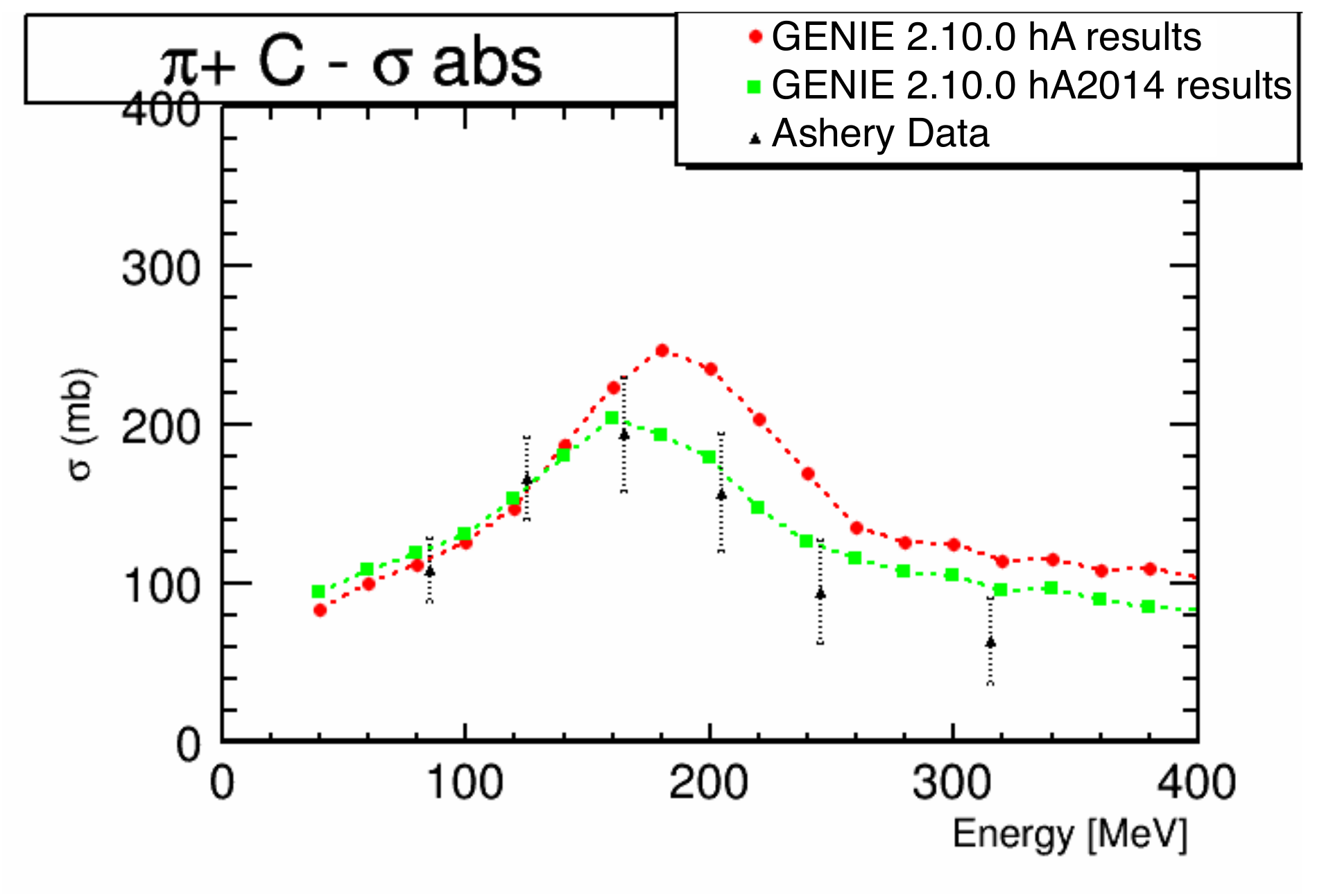}
  \includegraphics[height=0.25\textheight]{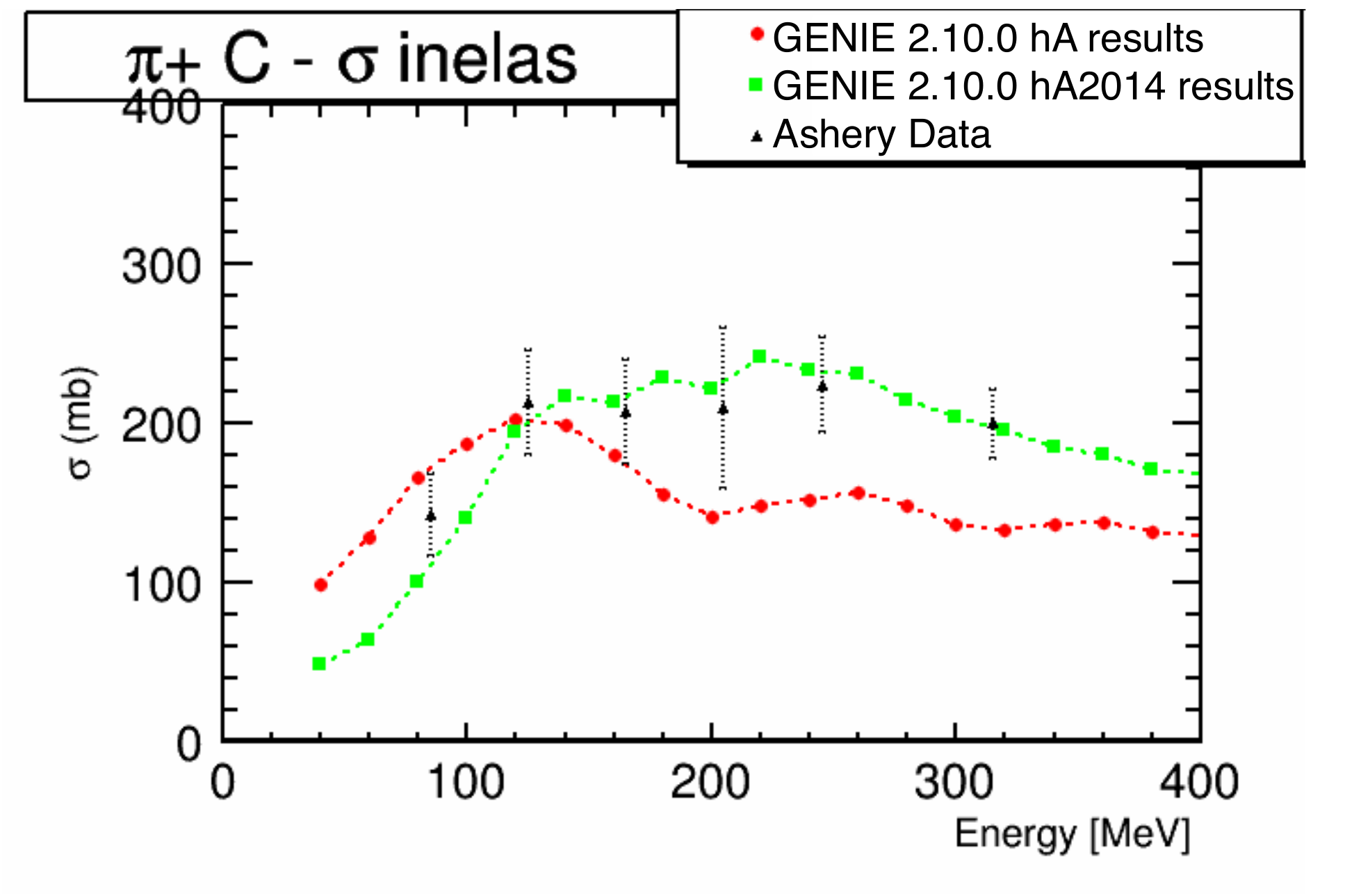}
  \caption{Comparison of new total cross sections for $\pi^+ C$ for new {\it hA2014} model with default model {\it hA}.}
  \label{fig:new_pi_total_xs}
\end{figure}

This release includes a new determination of the relative probabilities of pion interaction fates in the hA intranuclear cascade simulation.   
The default code for choice of final state channel uses data and model results, e.g. charge exchange vs. absorption, for a Fe target for all probes.  
These results are then unchanged for other nuclei which produces $\sim$ 20\% deviations from pion interaction data, which is much more available than for protons or neutrons.
The new alternate {\it hA2014} model includes a wide range of data for other nuclei for $\pi^\pm$ so that much less extrapolation is needed.
To enable it, set the parameter \texttt{HadronTransp-Model} to \texttt{genie::hAIntranuke2014/Default} where the default value is \texttt{genie::hAIntranuke/Default}.
The new data is mostly from Ashery (Li, C, Al, Fe, Nb, Bi)~\cite{Ashery:1981tq} but includes other sources~\cite{Navon:1983xj,Bowles:1981my,Lehmann:1998im,Aniol:1986kq,Brinkmoeller:1993si}.
To calculate the fractions for {\it hAIntranuke}, total cross sections~\cite{Clough:1974qt} and additional inputs are also needed.  
For higher energy pions (Ashery highest energy is 315 MeV), the Mashnik CEM03 Monte Carlo calculations for Fe ~\cite{Mashnik:2000up,Mashnik:2005ay} are still used. Other nuclei are simulated assuming $A^{\frac{2}{3}}$ scaling which is a good approximation when there is no data to use directly. 
At low energies, 0 and 50 MeV, the required values can be constructed from existing data~\cite{Saunders:1996ic,Navon:1983xj,Friedman:1991it}.
Here, the $\pi^-$ total cross sections should be larger than those for $\pi^+$.
Splines are built with almost all available data.  
Some data values cause sharp features in the splines.  
In those cases, individual data points were shifted within the estimated error to produce a smooth result.
Total inelastic cross section data from Ashery at 85 MeV is incompatible with the newer Aniol~\cite{Aniol:1986kq} data and was therefore not used.
Total absorption cross section data from Nakai~\cite{Nakai:1980cy} is not compatible with Ashery~\cite{Ashery:1981tq} and was not used.  
New results for $\pi^+ C$ are shown in Figure~\ref{fig:new_pi_total_xs} for the total absorption and inelastic cross sections.

\subsection{Cross Section Models}

 In 2.10.0 we made one significant change that impacts the global physics model - the GNU Scientific Library \cite{refgsl} is the new default numerical integrator. 
This change caused some minor numerical perturbations in the behavior of many of the cross section calculations.    A sign error for the $\Delta_s$ contribution 
to the axial form factor in neutrino-neutron NC elastic scattering has also been fixed, resulting in a roughly 25\% decrease in the cross section for this process.  

\subsubsection{High Energy Cross Sections}
In order to improve the accuracy of GENIE DIS calculations at energies greater than 100 GeV, the propagator terms for the W and Z bosons have been added to the DIS CC and NC differential cross section calculations.  Since this is clearly an improvement to the calculation, 
and since it has negligible impact on the CC cross section below 50 GeV, it is included in the default cross section calculation. 
 
\subsubsection{Berger-Sehgal Resonant Pion}
The Berger-Sehgal (BS)~\cite{Berger:2007rq} and Kuzmin-Lyubushkin-Naumov (KLN)~\cite{Kuzmin:2003ji} models for $N^*$ resonances are very similar 
to the default Rein-Sehgal~\cite{Rein:1980wg} model, but include the effects due to the muon mass.  
BS includes an extra diagram that is not found in KLN.
Much of the original code for the resonance couplings is untouched.  
The new models are enabled by changing the resonance (RES) model in \texttt{UserPhysicsOptions.xml} from 
\texttt{genie::ReinSehgalRESPXSec/Default} (default Rein-Sehgal model) to either
\texttt{genie::BergerSehgalRESPXSec2014/Default} (for BS) or
\texttt{genie::KuzminLyubushkinNaumovRESPXSec2014/Default} (for KLN).

Work in MiniBooNE collaboration also improved the form factors which have remained unchanged in the Rein-Sehgal resonance model \cite{boonetn260, Nowak:2009se}. 
These are also included with parameters (\texttt{minibooneGV} and \texttt{minibooneGA} for new vector and axial form factors) in \texttt{UserPhysicsOptions.xml}.

\begin{figure} [tp]
  \centering
  \includegraphics[height=0.25\textheight]{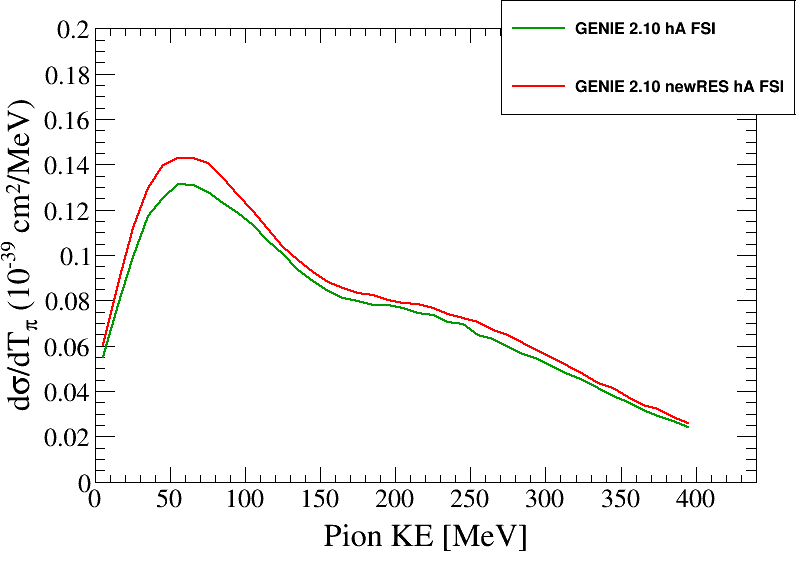}
  \includegraphics[height=0.25\textheight]{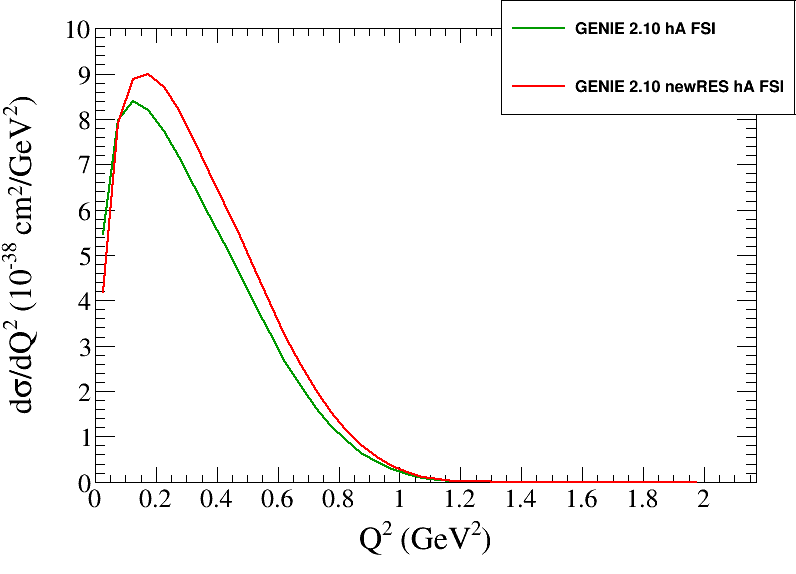}
  \caption{Comparison of new model (Berger-Sehgal with new form factors)
with default model.  }
  \label{fig:genie-BSnewGAGV}
\end{figure}

GENIE validations are shown
in Figure~\ref{fig:genie-BSnewGAGV} for 1 GeV $\nu_\mu C$ CC interactions.  
Distributions for true $Q^2$ and $\pi^+$ kinetic energy are supplied. 
Including the muon mass has the largest effect close to threshold.  
The differences in the plots are mostly due to changes in form factors.
 
\begin{figure}
  \centering
  \includegraphics[height=0.75\textheight]{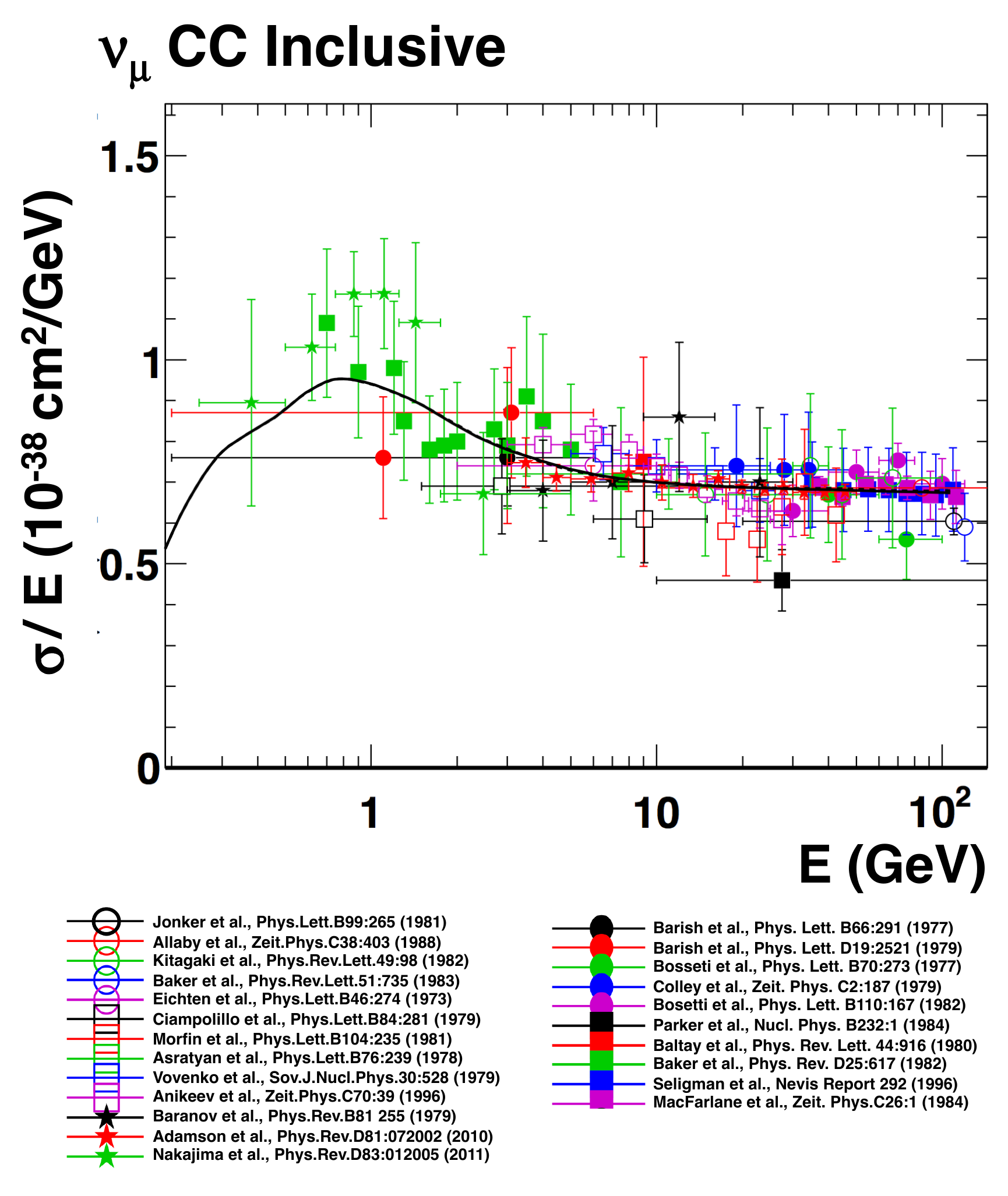}
  \caption{The GENIE 2.10.0 (solid black line), and 2.8.6 (dashed black line) muon neutrino inclusive CC cross section compared to data.}
  \label{fig:nu_ccinclusive}
\end{figure}

\begin{figure}
  \centering
  \includegraphics[height=0.75\textheight]{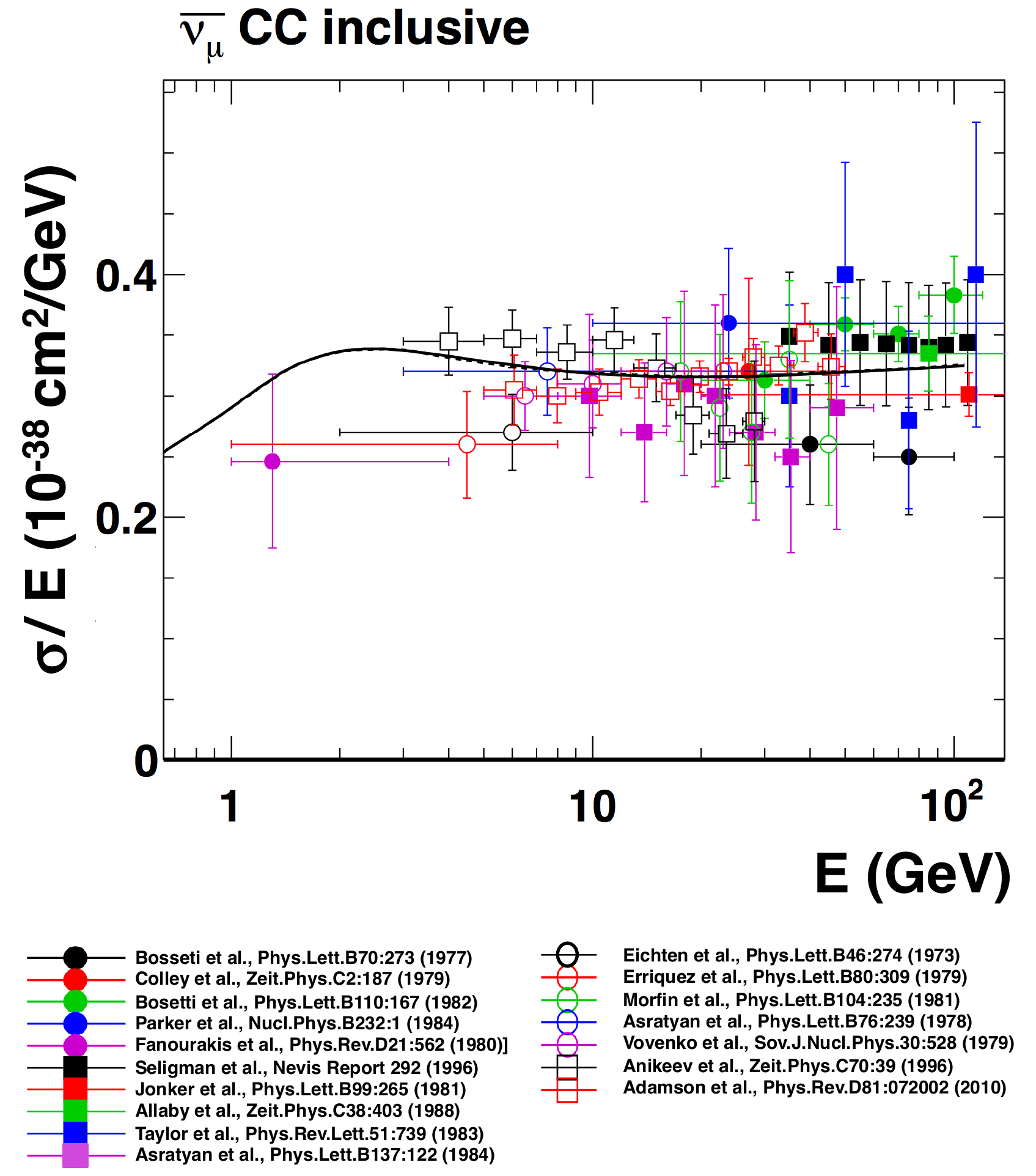}
  \caption{The GENIE 2.10.0 (solid black line), and 2.8.6 (dashed black line) muon anti-neutrino inclusive CC cross section compared to data.}
  \label{fig:nubar_ccinclusive}
\end{figure}

 Figures \ref{fig:nu_ccinclusive} and \ref{fig:nubar_ccinclusive} show the total charged-current cross section / energy for muon neutrinos and antineutrinos. 
Careful inspection of the total cross section shows some small changes resulting from the switch to gsl and the inclusion of the W/Z propagator 
terms, however, differences between the 2.10 and 2.8.6 cross sections are generally less than 1\% for neutrino energies less than 100 GeV.  

\section{New Interaction Models: Single Kaon Production}

There is a model for one entirely new interaction process in this release,  neutrino-production of single kaons from Athar et al. \cite{RafiAlam:2010kf}, generating events in the channels $\nu_l + p \rightarrow  l^- + K^{+} + p$,  $\nu_l + n \rightarrow  l^- + K^{0} + p$, and $\nu_l + n \rightarrow  l^- + K^{+} + n$.  This is the first $\Delta S=1$ process to be included in GENIE.    When running on a nuclear target, the process is embedded in the default nuclear model (for simulation of Fermi motion and intranuclear rescattering), in the same way that the free nucleon QEL model is incorporated into the nuclear model.   The cross section on a 
nucleus is taken as Z (N) times the free proton (neutron) cross section.  

\subsection{Single Kaon Production}

\begin{figure}
  \centering
  \includegraphics[width=0.95\textwidth]{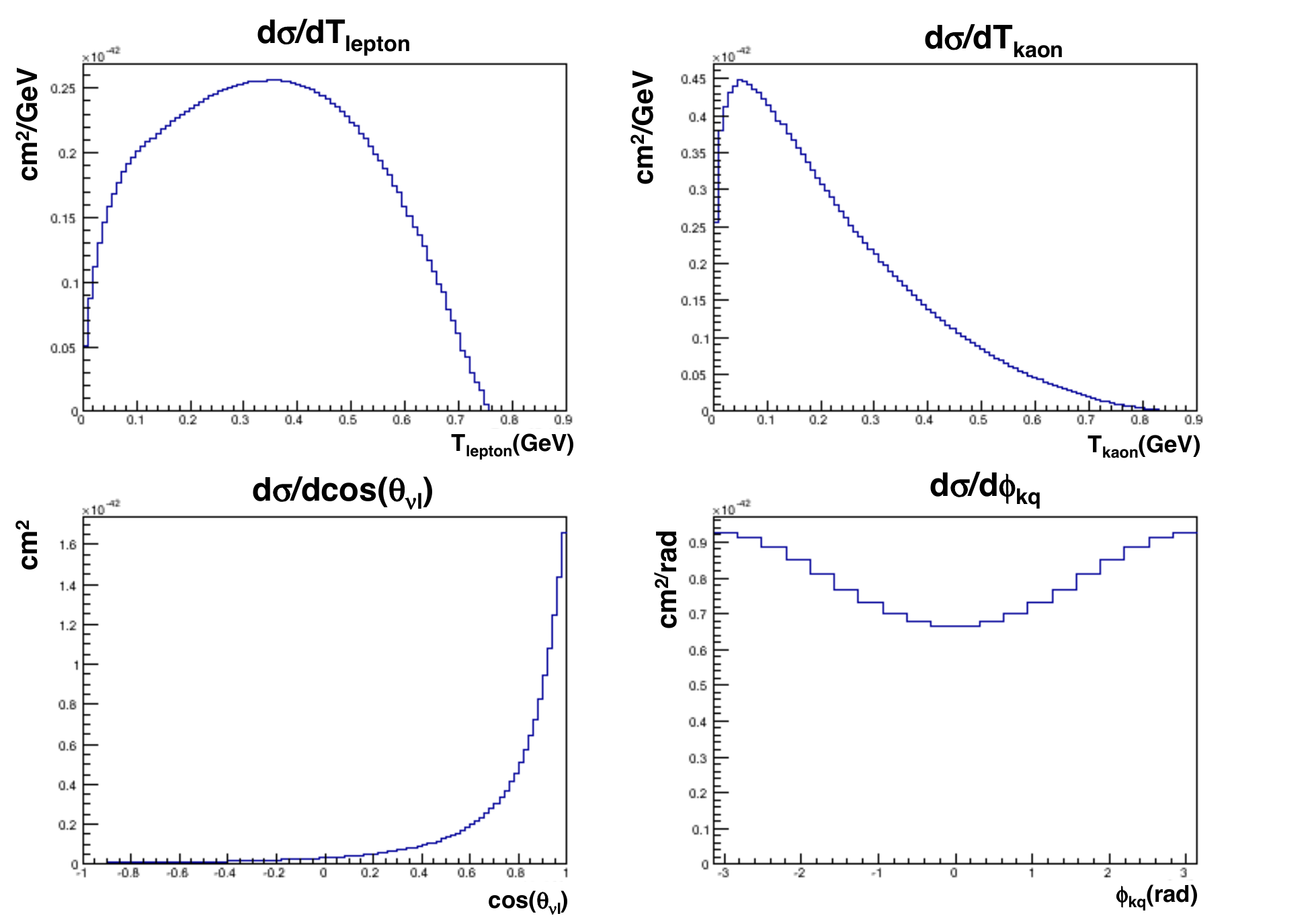}
  \caption{Kinematic distributions for the scattering of 1.5 GeV neutrinos in the channel  $\nu_\mu + p \rightarrow  \mu^- + K^{+} + p$.  \label{fig:singleton}  }
 \end{figure}

Figure~\ref{fig:singleton} shows the differential distributions produced in the scattering of 1.5 GeV muon neutrinos in the channel $\nu_\mu + p \rightarrow  \mu^- + K^{+} + p$, as calculated using a standalone C++ code that was validated against both the
original Fortran calculation, and the 2.10 GENIE implementation.  
In GENIE, this is implemented as a 4-fold differential cross section calculation in the outgoing lepton energy and scattering angle, the kaon energy, and $\phi_{Kq}$ (defined as the angle between the kaon-$\vec{q}$ plane and the lepton plane).
Handling integration of 4-fold differential cross sections with the necessary precision was one of the technical drivers of the transition to GSL that was part of this release.
The range of validity of this model is for neutrino energies up to 2 GeV (although the model will run for energies above that), and  no attempt has been made in this release to re-tune other processes which produce kaons, such as associated production.

This model is included as a specific implementation (in the GENIE code base as \\ {\tt AlamSimoAtharVacasSKPXSec2014}) of single-kaon production processes, which are identified by a new enum value ({\tt kScSingleKaon}) in the {\tt ScatteringType} object and related classes.
The {\tt DIS-CC-SINGLEK} event generator is currently configured to use this model as its default, and selecting the  {\tt SingleKaon} event generator list (i.e. as an input to gmkspl or gevgen), will enable this as the sole event generation thread.

\section{Tools and Experimental Interfaces}

We made a number of improvements and fixes to tools and interfaces distributed with GENIE. 
In particular, for this release, we made changes to the flux drivers, the reweighting machinery, and to some of the event generation applications.
Where possible, we provide code snippets to support the changes discussed in the text.

\subsection{Flux Drivers}

In order to provide better flexibility to users in applications that utilize 
flux drivers, three significant changes were made in this release:

\begin{enumerate}
\item
A {\tt GFluxDriverFactory}\ has been implemented; this allows flux drivers to 
self-register (Code~\ref{code:fdregister}) with the factory and be returned 
from the factory by providing the name as a string (Code~\ref{code:fdfactory}).
This will allow more applications to use different drivers interchangably
and for expansion of the available list by loading a library containing
as self-registering driver without need to rebuild the application.
\item
A common flux interface {\tt GFluxFileConfigI} (Code~\ref{code:ffconfig})
was introduced to unify configuration of flux drivers that depend on
external file sets, such as ntuple-based drivers.
These additions allow 
{\tt GNuMIFlux}, {\tt GSimpleNtpFlux} and 
the external {\tt GDk2NuFlux} to be used completely interchangeably with
any user configuration taking place via passed strings.  It also unifies
setting a limit on what neutrino flavors to return.  
For {\sc root}\ ntuple-based drivers it
unifies an interface for getting the underlying branch objects, allowing 
users to copy the corresponding detailed flux records to accompany a 
generated event.
Other flux drivers may start to incorporate these interfaces for additional
interchangeability.
\item
A second new common interface 
{\tt GFluxExposureI} (Code~\ref{code:fexposure}) 
unifies flux drivers that can report an ``exposure'' such as time or
protons-on-target.  Initially only
{\tt GNuMIFlux}, {\tt GSimpleNtpFlux} and 
the external {\tt GDk2NuFlux} have been migrated to use this; future
migration of others would expand the interchangablility.
\end{enumerate}

\begin{code}
\begin{Verbatim}[frame=single,fontsize=\footnotesize]
  // example registration of GDk2NuFlux with factory (from GDk2NuFlux.cxx)
  #include "Conventions/GVersion.h"
  #if __GENIE_RELEASE_CODE__ >= GRELCODE(2,9,0)
    #include "FluxDrivers/GFluxDriverFactory.h"
    // macro handling of namespace issues requires explicit split
    FLUXDRIVERREG4(genie,flux,GDk2NuFlux,genie::flux::GDk2NuFlux)
    // also exist FLUXDRIVERREG(afluxdriver)
    // and        FLUXDRIVERREG3(myns,myfluxdriver,myns::myfluxdriver)
  #endif 
\end{Verbatim}
\caption{Example code for registering new flux drivers with the factory}
\label{code:fdregister}
\end{code}

\begin{code}
\begin{Verbatim}[frame=single,fontsize=\footnotesize]
  genie::flux::GFluxDriverFactory& fdfactory =   
      genie::flux::GFluxDriverFactory::Instance();

  std::string fdname = "genie::flux::GSimpleNtpFlux"; // example name
  genie::GFluxI* myFluxDriver = fdfactory.GetFluxDriver(fdname);
  
  fdfactory.PrintConfig();
  GFluxDriverFactory has the following drivers registered:
    [  0]  genie::flux::GBartolAtmoFlux
    [  1]  genie::flux::GCylindTH1Flux
    [  2]  genie::flux::GFlukaAtmo3DFlux
    [  3]  genie::flux::GJPARCNuFlux
    [  4]  genie::flux::GMonoEnergeticFlux
    [  5]  genie::flux::GNuMIFlux
    [  6]  genie::flux::GSimpleNtpFlux
\end{Verbatim}
\caption{Example code for GFluxDriverFactory}
\label{code:fdfactory}
\end{code}

\begin{code}
\begin{Verbatim}[frame=single,fontsize=\footnotesize]
  genie::flux::GFluxFileConfigI* ffconfig = 
     dynamic_cast<genie::flux::GFluxFileConfigI*>(myFluxDriver);
  if ( ffconfig ) {
     // unified interface for some ntuple based flux drivers
     std::vector<std::string> filepatterns = ...
     std::string configstring = ...
     ffconfig->LoadBeamSimData(filepatterns,configstring);
     PDGCodeList pdglist = ...
     ffconfig->SetFluxParticles(pdglist);
     ffconfig->PrintConfig();
  }
\end{Verbatim}
\caption{Example code for GFluxFileConfigI interface}
\label{code:ffconfig}
\end{code}

\begin{code}
\begin{Verbatim}[frame=single,fontsize=\footnotesize]
  genie::flux::GFluxExposureI* expi = 
     dynamic_cast<genie::flux::GFluxExposureI*>(myFluxDriver);
  if ( expi ) {
     double        used = expi->GetTotalExposure();
     const char* eunits = expi->GetExposureUnits();
     double   probscale = myGMCJDriver->GlobalProbScale();
     double    exposure = used / probscale;
     std::cout << "Exposure:  " << exposure << " " << eunits << std::endl;
  }
\end{Verbatim}
\caption{Example code for GFluxExposureI interface}
\label{code:fexposure}
\end{code}

\subsection{Reweighting}

Two adjustments were made to the reweighting machinery: 

\begin{enumerate}
\item
In GENIE Release 2.8.2, the treatment of formation zones was improved and introduced separate parameters for formation times for mesons and nucleons.
The formation zone reweighing code has now been updated to reflect these changes.
\item
Prior to this release, the {\tt kRDcyTwkDial\_Theta\_Delta2Npi} knob in GENIE only affected the pion angular distribution for $\Delta(1232)^{++}$ events.
This has now been changed so that the reweighing is applied to all charge states of the $\Delta(1232)$.
\end{enumerate}

\subsection{Event Generation Applications}

The {\tt gevgen\_numi}\ application was renamed to {\tt gevgen\_fnal}\ to
indicate the wider applicability (\numi, \dune and booster based experiments).
The executable will dynamically pick up the external 
{\tt GDk2NuFlux}\ flux driver if available (i.e. there is no longer a build 
dependence of GENIE on {\tt Dk2Nu} for this feature).
For all supported (ntuple-based) beam related flux drivers the flux entry 
used from the input will be copied to a branch along side the  {\tt GHepRecord} 
and metadata from all input files will be copied to the output file.

\section{Technical Updates}

We made a small set of changes to the core ``technical'' components of GENIE. 
We define technical changes as those modifying the framework, framework interfaces, the event record, the configuration system, and the build system.

For GENIE 2.10.0 we restored the build Makefile structure from GENIE 2.8.0. 
In GENIE 2.8.2 we introduced a change that would stop the build on any error.
We have returned to the 2.8.0 behavior of continuing to attempt to build libraries even after one library has failed.

\subsection{Event Record Updates}

Three changes were made to the GENIE event record, enumerated below. 
Please note that this change is not backwards compatible in the sense that events produced with GENIE 2.10 will not be successfully read by older versions of GENIE.
Because GENIE uses ROOT for persistency, it should be possible to read event produced by older versions of GENIE. 

\begin{enumerate}
\item
An additional value has been added to the enumeration in the {\tt ScatteringType} class to identify single kaon events, as described above.
\item
The \texttt{XClsTag} object has been updated to include information needed to tag $\Delta S=1$ events, which can now be accessed via the {\tt isStrangeEvent}
and {\tt StrangeHadronPdg} methods.
\item
Reintroduced the {\tt DiffXSecVars}\ method which allows a query of the 
{\tt GHepRecord}\ for the current {\tt KinePhaseSpace\_t}\ value.
\end{enumerate}

\section{Conclusions}

While the default physics in GENIE 2.10.0 is largely the same as the last production series, GENIE 2.8, this release has made some important technical changes and introduced a set of new models for advanced users.

\section{Acknowledgements}

We are grateful to many members of the neutrino physics community who provided feedback on development versions of this release.
In particular, we would like to thank: I. Kagorin for comments on the KLN resonance model, R. Gran and J. Schwehr for a careful review of some numerical routines, T. Le for discovering a mistake in the Delta re-weighting code, and L. Alvarez-Ruso for finding a small mistake in the neutral current elastic scattering model.  

This work was supported by the Fermi National Accelerator Laboratory, operated by Fermi Research Alliance, LLC under Contract No. De-AC02-07CH11359 with the United States Department of Energy.  Work at Tufts University was supported by Department of Energy
grant DE-SC0007866.  
Work at the University of Pittsburgh was supported by Department of Energy grant DE-SC0007914.


\bibliographystyle{h-physrev}


\end{document}

%% file: preamble.tex
\usepackage{amsmath}
\usepackage{amssymb}
\usepackage{amsthm} 
\usepackage{bm} 
\usepackage{xspace}
\usepackage{upgreek}

\usepackage{lmodern}

\newcommand{\be}{\begin{equation}}
\newcommand{\ee}{\end{equation}}
\newcommand{\bea}{\begin{eqnarray}}
\newcommand{\eea}{\end{eqnarray}}

\numberwithin{lemma}{theorem}

\usepackage{moreverb} 

\newcommand{\Aligarh}{Department of Physics, Aligarh Muslim University, Aligarh-202 002, India}
\newcommand{\UBern}{University of Bern, Albert Einstein Center for Fundamental Physics, Laboratory for High Energy Physics (LHEP), Bern, Switzerland}
\newcommand{\FNAL}{Fermi National Accelerator Laboratory, Batavia, Illinois 60510, USA}
\newcommand{\Hampton}{ Dept. of Physics, Hampton University, Hampton, VA 23668, USA}
\newcommand{\Lancaster}{ Department of Physics, Lancaster University, Lancaster, LA0 4YW, United Kingdom}
\newcommand{\Liverpool}{Department of Physics, University of Liverpool,  Liverpool, L69 7ZE, UK}
\newcommand{\MSU}{Department of Physics and Astronomy, Michigan State University, East Lansing, MI 48824, USA}
\newcommand{\Pittsburgh}{Department of Physics and Astronomy, University of Pittsburgh, Pittsburgh, Pennsylvania 15260, USA}
\newcommand{\Rochester}{Department of Physics, University of Rochester, Rochester, New York 14610 USA}
\newcommand{\STFC}{Particle Physics Department, STFC Rutherford Appleton Laboratory, Harwell Oxford, OX11 0QX, UK}
\newcommand{\Tufts}{Department of Physics and Astronomy, Tufts University, Medford, Massachusetts 02155, USA}
\newcommand{\Warwick}{Department of Physics, University of Warwick, Gibbet Hill Road, Coventry CV4 7AL, UK}
\newcommand{\WM}{Department of Physics, College of William \& Mary, Williamsburg, Virginia 23187, USA}
\newcommand{\Wisconsin}{Dept. of Physics and Wisconsin IceCube Particle Astrophysics Center, University of Wisconsin, Madison, Wisconsin 53706, USA}

\newcommand{\dune}{DUNE\xspace}
\newcommand{\numi}{NuMI\xspace}

\newcommand{\pythia}{PYTHIA\xspace}

\usepackage{fancyvrb}

\usepackage{caption} 
\usepackage{newfloat}
\DeclareFloatingEnvironment[fileext=snip,listname={List of Code Snippets},name=Code,placement=tbphH]{code}
